\documentclass[11pt]{iopart}
\usepackage{graphicx}
\usepackage{subfigure}
\usepackage{morefloats}
\usepackage{color}
\usepackage{setspace}
\usepackage{floatrow}
\usepackage{amsmath,bm}
\usepackage[pagewise]{lineno}
\newif\ifproofread
\newcommand{\markchange}[1]{
\ifproofread
{\color{red}#1}
\else
#1
\fi
}

\begin{document}
%Change this to turn on/off coloring of changes
\proofreadfalse

\title[]{Fluctuation characteristics of the TCV snowflake divertor measured with high speed visible imaging}
\author{ N. R. Walkden$^{1}$, B. Labit$^{2}$, H. Reimerdes$^{2}$, J. Harrison$^{1}$,  T. Farley$^{1,3}$, P. Innocente$^{4}$, F. Militello$^{1}$, the TCV Team\footnote{See the author list of "S.Coda et al., Nucl. Fusion 57 (2017) 102011"} and the MST1 Team\footnote{See the author list of "H.Meyer et al., Nucl. Fusion 57 (2017) 102014"}
        \\ \small{$^{1}$ CCFE/UKAEA, Culham Science Centre, Abingdon, Oxfordshire, OX14 3DB, UK}
        \\ \small{$^{2}$ Swiss Plasma Center, Ecole Polytechnique Federale de Lausanne, Station 13, CH-1015, Lausanne, Switzerland}
        \\ \small{$^{3}$ Department of Electrical Engineering and Electronics, Univ. Liverpool, L69 3GJ, UK}
        \\ \small{$^{4}$ Consorzio RFX, Corso Stati Uniti, 4 - 35127, Padova, Italy}
        \\ Email: \texttt{nick.walkden@ukaea.uk} }
\date{}

\begin{abstract}
Tangentially viewing fast camera footage of the low-field side snowflake minus divertor in TCV is analysed across a four point scan in which the proximity of the two X-points is varied systematically. The motion of structures observed in the post-processed movie shows two distinct regions of the camera frame exhibiting differing patterns. One type of motion in the outer scrape-off layer remains present throughout the scan whilst the other, apparent in the inner scrape-off layer between the two nulls, becomes increasingly significant as the X-points contract towards one another. The spatial structure of the fluctuations in both regions is shown to conform to the equilibrium magnetic field. When the X-point gap is wide the fluctuations measured in the region between the X-points show a similar structure to the fluctuations observed above the null region, remaining coherent for multiple toroidal turns of the magnetic field and indicating a physical connectivity of the fluctuations between the upstream and downstream regions. When the X-point gap is small the fluctuations in the inner scrape-off layer between the nulls are decorrelated from fluctuations upstream, indicating local production of filamentary structures. The motion of filaments in the inter-null region differs, with filaments showing a dominantly poloidal motion along magnetic flux surfaces when the X-point gap is large, compared to a dominantly radial motion across flux-surfaces when the gap is small. This demonstrates an enhancement to cross-field tranport between the nulls of the TCV low-field-side snowflake minus when the gap between the nulls is small.
\end{abstract}

\section{Introduction}
Optimising divertor conditions for a tokamak based fusion reactor is a key area of research activity\cite{ITER2007CH4,LipschultzNF2007,WenningerNF2015}. The high heat fluxes expected to impinge on the surface of the divertor target must be mitigated to ensure machine survival. This places a degree of importance on heat transport contributing to the profile of the heat flux to the material surface \cite{EichPRL2011}. Likewise, particle transport has an important role in aspects of the machine operation including fuel retention, material migration, erosion and sputtering \cite{BrezinsekNF2015,BehrischJNM2003}. Moreover, due to the challenging heat fluxes the standard operating scenario for ITER will likely include a partially detached divertor \cite{EichNF2013}, which greatly reduces the heat flux to the divertor target. Detachment onset and detachment front control are very sensitive to both the electron temperature and the electron density along the divertor leg which in turn are sensitive to the transport processes within that region. Transport processes parallel to the magnetic field are routinely captured in two-dimensional fluid codes such as SOLPS \cite{SOLPS}, EDGE2D \cite{EDGE2D} or UEDGE \cite{UEDGE} and may be considered reasonably well understood. Perpendicular transport processes on the other hand are generally rather poorly understood, and are usually captured heuristically in 2D fluid codes. Upstream, adjacent to the core plasma, perpendicular transport in the scrape-off layer is intermittent \cite{AntarPoP2003,BoedoPoP2003,GarciaNF2007,DudsonPPCF2005,XuNF2009} and highly non-diffusive \cite{NaulinJNM2007,GarciaJNM2007}. A significant component of the heat and particles carried into and through the SOL perpendicular to the magnetic field is carried in intermittent coherent turbulent objects known as filaments/blobs \cite{BoedoPoP2003,KrashenninikovPLA2001,D'IppolitoReview}. These structures have been documented and analysed on many machines worldwide and are routinely modelled in both isolated filament simulations \cite{YuPoP2003,GarciaPPCF2006,YuPoP2006,WalkdenPPCF2013,EasyPoP2014} as well as fully turbulent simulations. Filaments are also present below the X-point. Recent analysis of the MAST divertor region \cite{HarrisonJNM2015,HarrisonPoP2015,WalkdenNF2017} showed a rather complex multi-region picture of cross-field transport. Turbulent structures appear in the far-SOL of the outer divertor leg, which connect to structures born upstream which flow down into the divertor volume via parallel transport. Filaments also appear in the private-flux region (PFR) which are born in the inner divertor leg \cite{HarrisonPoP2015}, and in the near-separatrix region of the outer divertor leg. These divertor localised outer leg filaments exist with shorter lifetimes than elsewhere. Local to the X-point there are not detectable fluctuations when measured with tangential view high speed imaging, leading to the definition of the quiescent X-point region (QXR) \cite{WalkdenNF2017}. The QXR conforms well to the magnetic flux-surfaces local to the X-point. The presence of the QXR indicates that the geometry of the null region can have a significant impact on the turbulent transport processes in the divertor volume that contribute to profile structures at the divertor surfaces. 
\\To complement the conventional divertor design a suite of 'alternative' divertor concepts exist which exploit novel designs to optimize conditions at the divertor surface towards tolerable levels in a future tokamak based fusion reactor. One such advanced divertor concept is the 'snowflake' divertor \cite{RyutovPoP2007,PirasPPCF2009}. In the ideal snowflake divertor a second-order null in the poloidal magnetic field is achieved such that both the conditions $B_{p} = 0$ and $\nabla B_{p} = 0$ are satisfied \cite{RyutovPoP2007}. This leads to a locally hexagonal structure in the poloidal field, with two limbs forming the separatrix that encircles the core plasma, whilst the other four connect to targets at the divertor forming the primary plasma-surface interface of the machine. The snowflake has a larger region of low poloidal magnetic field by virtue of its higher order null point compared to the standard divertor. This leads to strong flaring of the magnetic flux surfaces local to the null region which increases the plasma volume in the divertor available to radiate power \cite{ReimerdesNF2017}. A drastic increase in field-line connection length is also present which leads to enhanced power losses along the magnetic field-line compared to a conventional divertor. 
\\In practice the ideal snowflake configuration is a single point in the operational space of the machine and requires unobtainable precision in the control system of the machine to maintain. Rather two alternate configurations, the snowflake minus (SF-) and snowflake plus (SF+), are formed by bringing two X-points into close proximity \cite{LabitNME2017}. The primary route by which a reduced net power to the divertor target is achieved in the Snowflake divertor is by redistributing power and particles from upstream onto the two divertor legs that are not topologically connected to the upstream SOL \cite{VijversNF2014}. This relies on cross-field transport in some fashion to provide a mechanism by which this redistribution can occur. As has been shown on MAST \cite{WalkdenNF2017}, the null region can impact the nature of turbulent cross-field transport and even inhibit it entirely. In the snowflake divertor however, the expanded region of low poloidal magnetic field has been predicted to give rise to a 'churning' mode providing plasma convection to the snowflake divertor legs that are not primarilly connected to the plasma core \cite{RyutovPS2014}. The presence of a second X-point in the outboard side SOL has been shown to impact upstream SOL profiles \cite{TsuiAPS2016} leading to moderate increases in the near SOL density and electron temperature. This paper now seeks to provide a thorough characterisation of the properties of scrape-off layer fluctuations near the null region in snowflake plasmas with an X-point in the outer SOL. 
\\In section \ref{Sec:Exp} the setup of the fast camera and the experimental data used in this study are introduced. In section \ref{Sec:Results} the main results of the study are described. General properties of the fluctuations within the movie time-series are analysed, before a detail analysis of the spatial and temporal characteristics of fluctuations in different regions of the movie are carried out. Section \ref{Sec:Disc} discusses the results and presents a hypothesis to describe the observed fluctuation behaviour as the X-point gap narrows and section \ref{Sec:Conc} concludes.

\section{Experimental Setup}
\label{Sec:Exp}
\begin{figure}[htbp]
    \includegraphics[width=0.7\textwidth]{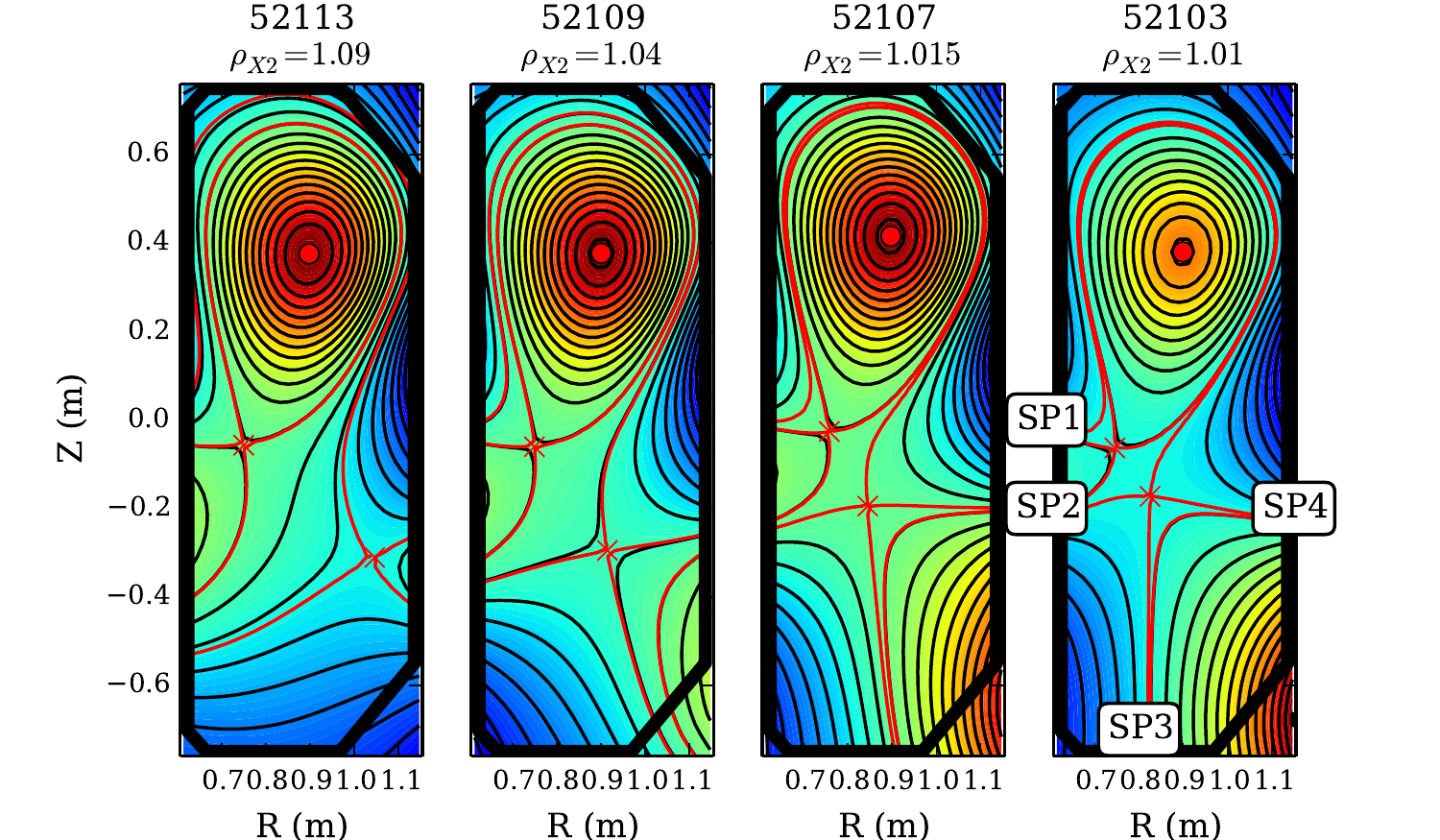}
    \caption{Magnetic equilibria in the four plasmas studied in this contribution representing a scan in the parameter $\rho_{X2}$, calculated at $t = 0.91$s. All equilibria are calculated at a time of 0.91s at a comparable density. The individual divertor leg strike points are labelled SP1 to SP4 with SP1 as the inner upper strike point and the labelling proceeding anti-clockwise. This is displayed for the case of 52103, and the convention is the same for all four. Color contours in this figure, as well as all other figures illustrating the magnetic equilibrium represent the variation of the poloidal magnetic flux.}
    \label{Fig:Equilibria}
\end{figure}

\begin{figure}[htbp]
    \includegraphics[width=0.7\textwidth]{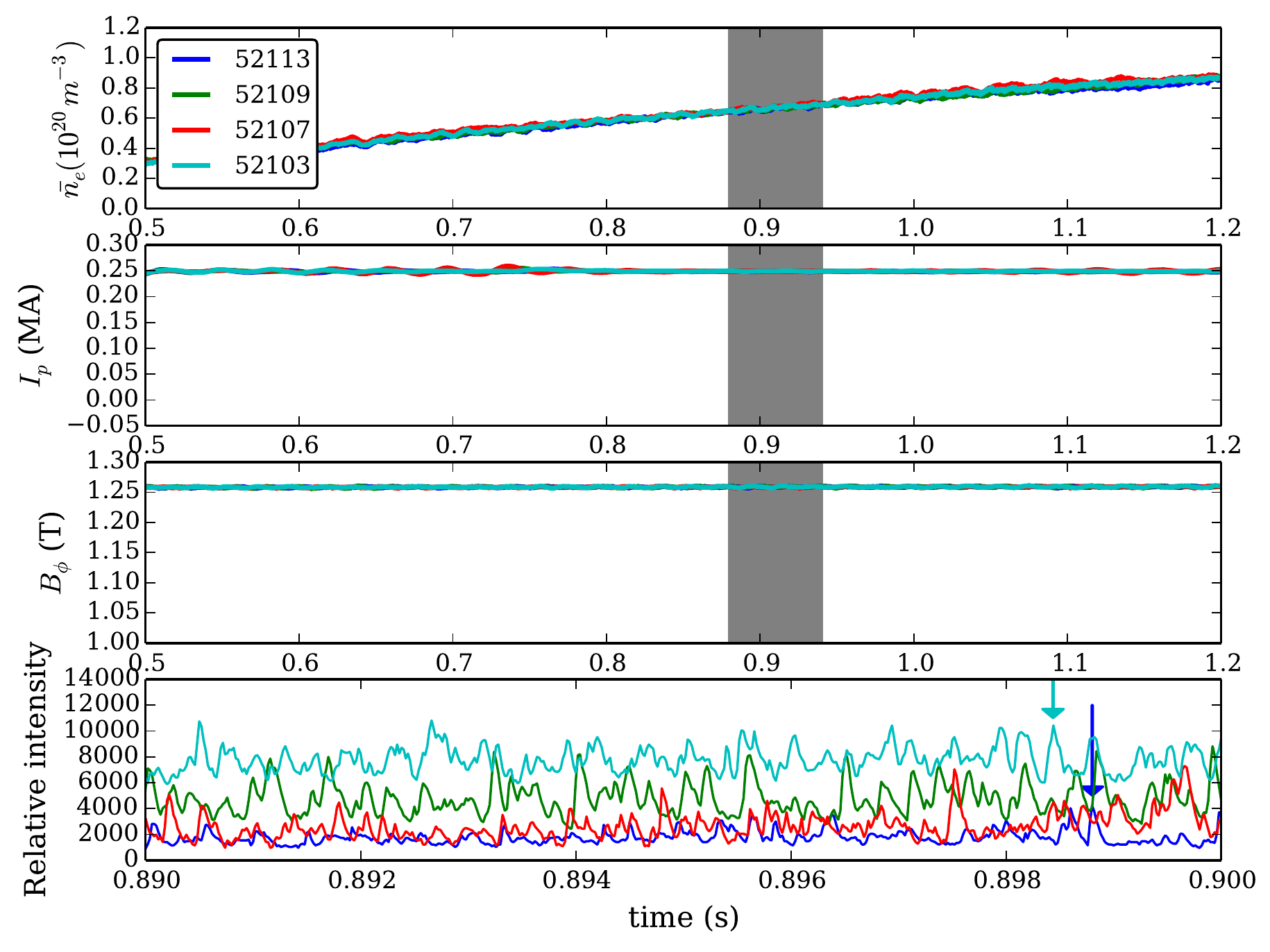}
    \caption{First three rows: Line-averaged density, plasma current and toroidal magnetic field for the four plasmas in the $\rho_{X2}$ scan. The window within which analysis has been conducted is shaded in grey. Lowest row: Time-trace of the raw pixel intensity measured on a camera pixel that views the outer scrape-off layer. Two specific fluctuations which can be cross-referenced with figure \ref{Fig:Cam_examples} are indicated with arrows.}
    \label{Fig:Signal_traces}
\end{figure}

\begin{figure}[htbp]
    \includegraphics[width=0.5\textwidth]{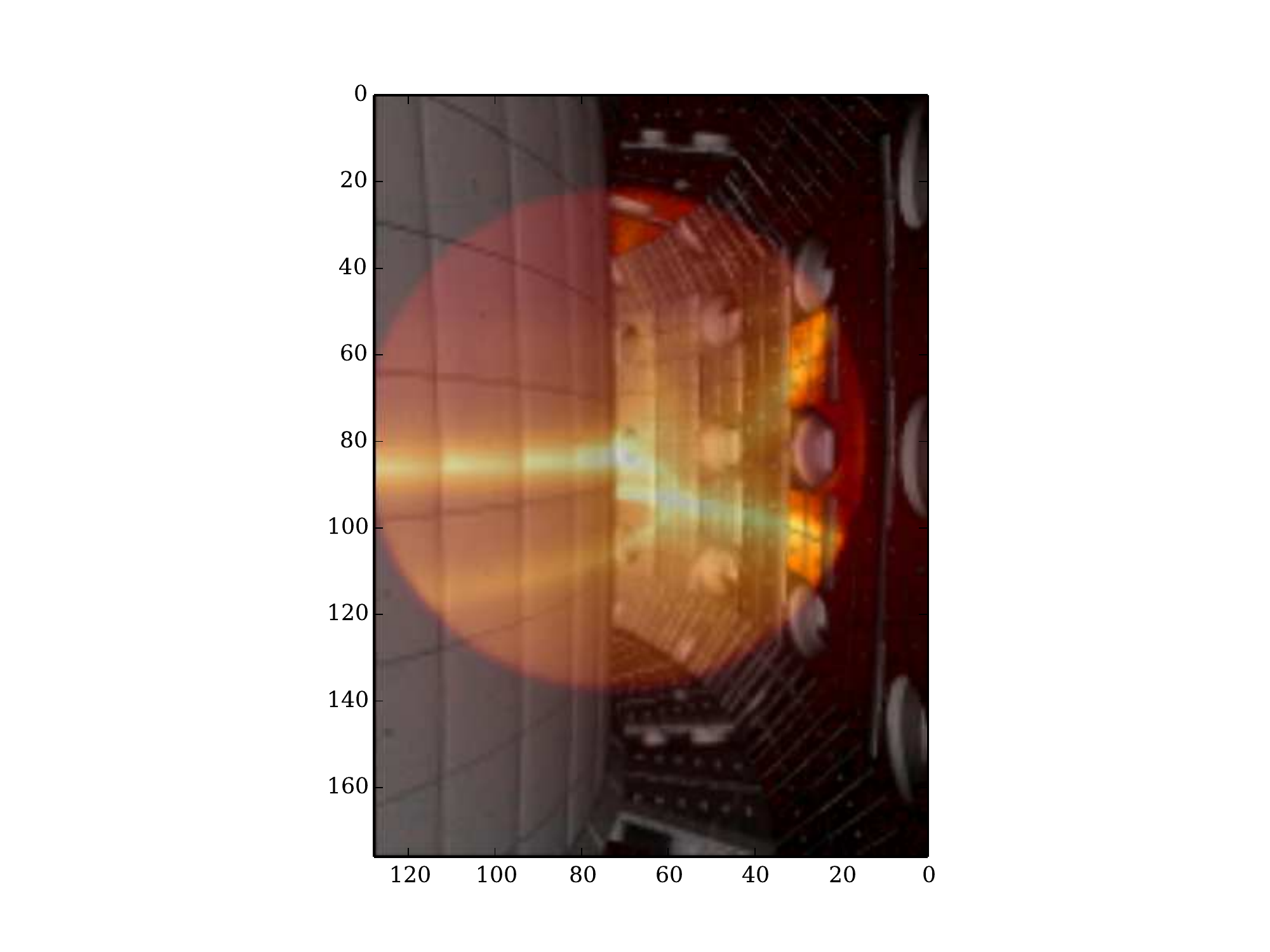}
    \caption{Rendering of the camera view into a CAD visualisation of the TCV vessel. Overlaid in a false heat-map is an image from 52103 to show the context of the camera view.}
    \label{Fig:CAD_rendering}
\end{figure}

\begin{figure}[htbp]
    \includegraphics[width=0.4\textwidth]{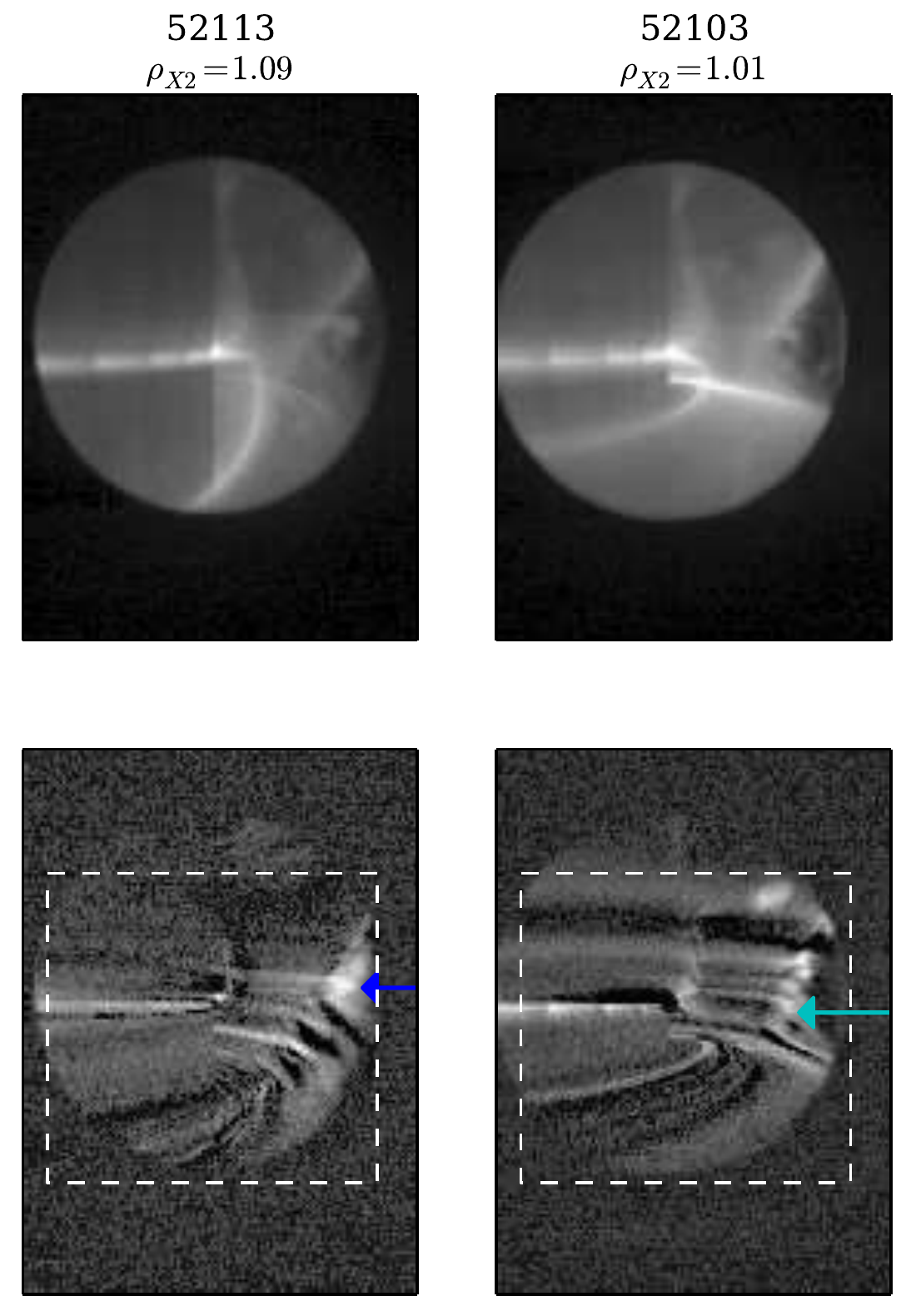}
    \caption{Example of the raw movie \markchange{(upper row) and post-processed movies (lower row)} in plasmas 52113 and 52103. The same \markchange{post-processing} technique is applied to the other two plasmas in the scan. A gamma enhancement with $\gamma = 0.5$ has been applied to the images shown here for visual clarity, but is not used in the subsequent analysis. Since a significant proportion of the camera sensor is dark, a region of interest (ROI) has been chosen that encompasses the area of the movie where the \markchange{magnetic} null region is visible. This is the region inside the dashed box in the lower row.\markchange{ Arrows in the lower row of figures point to pixels on which fluctuations shown in figure \ref{Fig:Signal_traces} are measured}.}
    \label{Fig:Cam_examples}
\end{figure}

The experiments studied here were performed on TCV \cite{CodaNF2015} and comprise a four-point scan in the parameter $\rho_{X2}$ which parameterises the distance between two X-points in the LFS SF- configuration. $\rho_{X2}$, as defined in ref \cite{LabitNME2017}, is given by
\begin{equation}
    \rho_{X2} = \sqrt{\frac{\psi_{X2} - \psi_{0}}{\psi_{X1} - \psi_{0}}}
\end{equation}
where $\psi_{X2},\psi_{X1}$ and $\psi_{0}$ are the poloidal magnetic flux at the secondary X-point, primary X-point and magnetic axis respectively such that $\rho_{X2}^{2}$ gives the normalised poloidal flux of the flux-surface that intersects the outer X-point. The two topological variants of the snowflake, the snowflake minus (SF-) and snowflake plus (SF+) are described by $\rho_{X2} > 1$ and $\rho_{X2} < 1$ respectively. In this contribution the low-field side (LFS) SF- configuration is studied, where the secondary X-point is situated in the LFS SOL. At $\rho_{X2} = 1.09$ the outer separatrix terminates on the outer wall and the X-points are strongly separated from one-another. At the opposite end of the scan, at $\rho_{X2} = 1.01$, the two separatrices are very close to one-another upstream and the X-points are in close proximity in the divertor. Figure \ref{Fig:Equilibria} shows the magnetic configuration for the four plasmas in the $\rho_{X2}$ scan.  
The strike points in the snowflake are labelled as SP1 to SP4 sequentially, beginning at the inner-upper strike point and proceeding anti-clockwise. These have been labelled in figure \ref{Fig:Equilibria} for plasma 52103, but the convention is the same for all four cases. \markchange{This allows for the definition of the ISOL (inner scrape-off layer) and OSOL (outer scrape-off layer). The ISOL is the region of the SOL that spans the two nulls and connects the upstream low-field side SOL through to SP2, whilst the OSOL is the region outside of the secondary null in the low-field side SOL that connects from upstream to SP4. Here upstream refers to the region of the SOL that is adjacent to the core plasma, above the null region, whilst downstream refers to the region below the nulls. These terms are used here purely as a way of differentiating the regions inside and outside the secondary separatrix and do not relate to other definitions that may be found in literature. These are not rigorous definitions and are used here only for ease of labelling the different characteristics of the plasma in different regions.} The scan is conducted in Ohmic L-mode attached Deuterium plasmas with no external heating, with a slow fuelling ramp leading to a gradual increase in line-averaged density throughout the shot. \markchange{Low density Ohmic L-mode is the selected scenario here because it is easy to control and repeatable. Higher power and/or higher fuelling rates bring the added complication of L-H transitions and/or detachment, which have not been well characterised for snowflake plasmas. Operating in low density Ohmic L-mode allows this study to be carried out cleanly. Extension to other operating regimes may be a good topic for future research.}
\\Other than the proximity of the secondary X-point, all plasma parameters are comparable during the scan. \markchange{Introducing a second X-point into the outboard SOL, as is done here in LFS SF- plasmas, has been shown to have a very mild effect on upstream scrape-off layer properties \cite{TsuiAPS2016,RobertoCommun} with a slight increase in pressure shown in the very near SOL, but nothing that may indicate a strong change in turbulence drive present in that region.} Time windows for analysis are chosen such that the plasma density in each case is comparable and the plasmas are in an attached regime. Figure \ref{Fig:Signal_traces} shows the line-averaged density, plasma current and toroidal magnetic field evolution for the four equilibria in the $\rho_{X2}$ scan.
\markchange{In the final row of figure \ref{Fig:Signal_traces}, time-traces are shown from a camera pixel in the OSOL region for each plasma. The signals fluctuate strongly, and these fluctuations correspond to filamentary structures. To illustrate this, two specific fluctuations have been indicated which correspond to structures shown in figure \ref{Fig:Cam_examples}.}
\\The camera model used for analysis here is a Photron APX-RS and was operated at a frame-rate of 50kHz with an integration time of $20\mu$s and a pixel-resolution of 128x176 in the horizontal and vertical dimension respectively. At the camera tangency angle this provides a spatial resolution of $\approx 5$mm. The camera was mounted to a midplane viewing port on the TCV vessel and had a tangential line of sight towards the plasma, encompassing approximately a half-view of the interior. Figure \ref{Fig:CAD_rendering} shows a rendering of the camera view into the TCV vessel, overlaid with a false color image from plasma 52103.
The camera is unfiltered and sensitive to the visible spectrum meaning that any emission from the plasma at visible wavelengths contributes to the camera image. Since the intensity of the camera image is uncalibrated and the image is unfiltered no attempt has been made to characterise the spectral properties of the emission. Scrape-off layer fluctuations present in the camera image are likely to be caused primarily by fluctuations in the plasma conditions which can be inferred by their adherence to the structure of the background magnetic field. Also shown in the final row of figure \ref{Fig:Signal_traces} are typical time series of the raw intensity measured on a pixel of the camera sensor viewing the OSOL region of the plasma during the analysis time window. In each case the signal is observed to fluctuate strongly, \markchange{and displays similar characteristics to} signals obtained on Langmuir probes \markchange{and gas-puff imaging diagostics on many other machines \cite{D'IppolitoReview}}. To isolate the fluctuating component of the movie a background subtraction technique has been applied where the pixel-wise minimum of the intensity of the frame of interest alongside the ten preceding frames is subtracted from the frame of interested \markchange{such that 
\begin{equation}
    D_{I}(t,i,j) = I(t,i,j) - \min\left[I(t - k,i,j)\right], (k = 0,1,2,...,10)
\end{equation}
where $D_I(t,i,j)$ and $I(t,i,j)$ are the background subtracted and raw pixel count on pixel $i,j$ at time index $t$ respectively}. This removes the slow varying background emission and isolates only the rapidly fluctuating component of the light emission measured by the camera. The technique has been successfully applied to data from MAST for the analysis of fluctuations near the midplane \cite{AyedPPCF2009,DudsonPPCF2008,KirkPPCF2016} and in the divertor \cite{HarrisonJNM2015,HarrisonPoP2015,WalkdenNF2017}. \markchange{Alternative background subtraction techniques were tested (for example a mean based background subtractor as opposed to the minimum used here) with no qualitative or quantitative impact on results presented within this paper. The minimum based subtractor was chosen here since SOL fluctuations tend to be positive perturbations on top of a background plasma \cite{D'IppolitoReview}, indicating that the background may be well extracted with a minimum based technique.} In figure \ref{Fig:Cam_examples} examples of movie frames from TCV plasma discharges 52113 and 52103 are shown before and after post-processing, demonstrating the effective isolation of the fluctuating component of the light. \markchange{ The arrows shown in the lower rows indicate the position in the image plane where the fluctuations shown in figure \ref{Fig:Signal_traces} are measured, and in particular where the peaks indicated in the lower row of figure \ref{Fig:Signal_traces} occur.  This shows that fluctuations in the pixel count can be associated with filamentary fluctuations.}
\markchange{From figure \ref{Fig:Cam_examples}, it is possible to identify SP's 1,2 and 4 in the raw movie from plasma 52103, where the X-point gap is small. In 52113 SP4 in the poloidal plane,is outside the camera field of view and can only weakly be distinguished in the far-field of the image, whilst SP's 1 and 2 are clearly visible. Since the strike point emission is relatively slowly varying, the background subtraction is effective in removing it from the analysed data, with the exception of some frames in 52103 where the emission from SP1 saturates the camera pixel. This does not impact the measurements made.}
\\The analysis conducted in the following sections is taken over a time-series of 3000 frames ($60$ms) in a time-range where the plasma conditions are relatively stationary and comparable between the different plasmas. The analysis window is highlighted with a shaded area in  figure \ref{Fig:Signal_traces}. A large time-series is required to ensure accuracy of the various statistical measures used in the forthcoming analysis. $60$ms was found to provide a good balance between the need for stationary plasma conditions and statistical convergence of the measures used for analysis. 

\section{Results}
\label{Sec:Results}
\begin{figure}[htbp]
    \includegraphics[width=\textwidth]{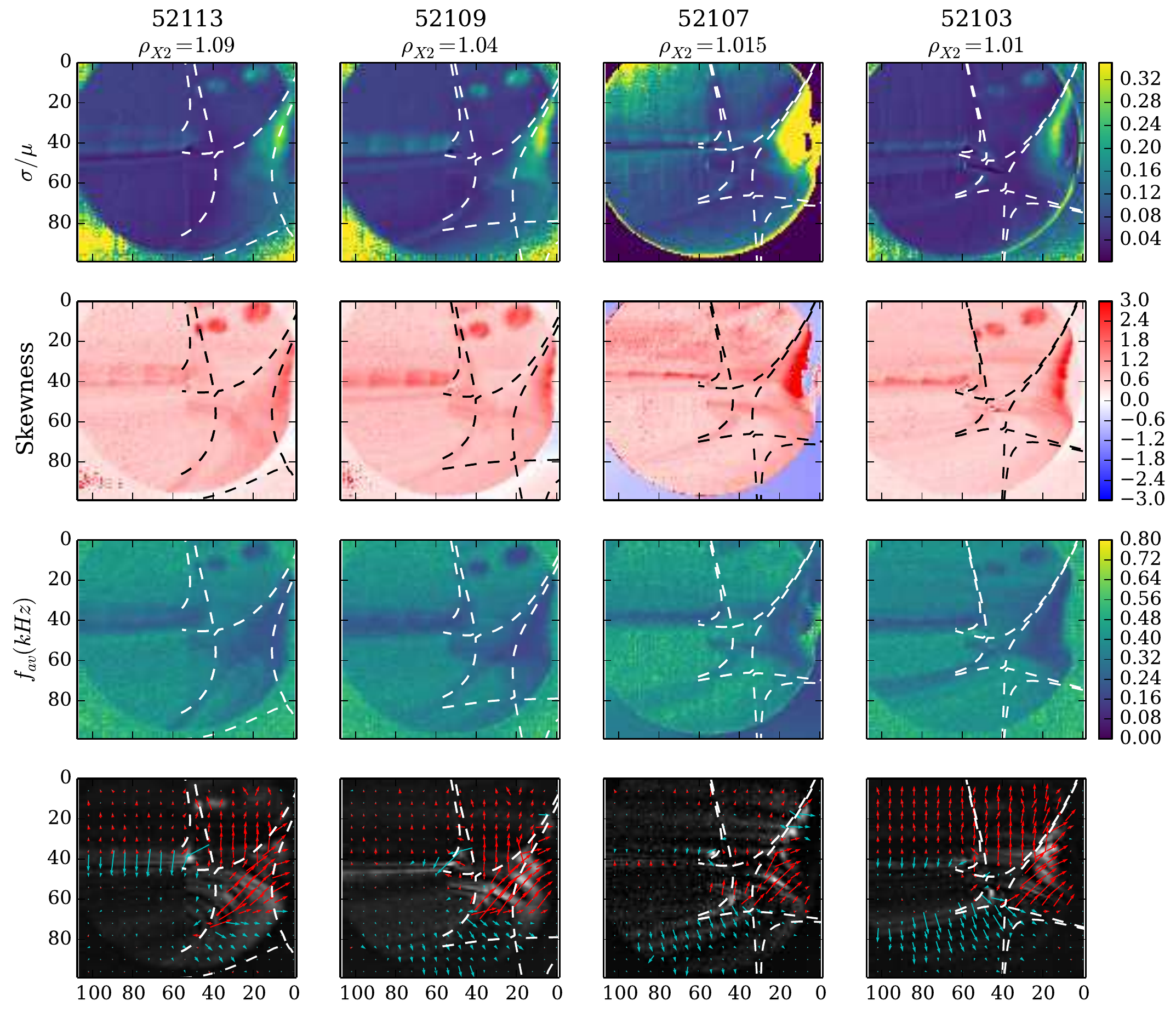}
    \caption{\markchange{Pixel-wise fluctuation amplitude, skewness, average frequency and apparent flow for each plasma in the $\rho_{X2}$ scan}. Separatrices corresponding to the two X-point flux-surfaces are projected onto the image.}
    \label{Fig:Fluct_amps_flow}
\end{figure}

In all cases within the scan of $\rho_{X2}$ a strong fluctuating component of the light viewed by the camera is present in the SOL. In figure \ref{Fig:Fluct_amps_flow} \markchange{four measurements of the pixel count from the movie have been made for each case in the $\rho_{X2}$ scan for each pixel in the camera frame (i.e pixel-wise). The fluctuation amplitude, given by the mean $\mu$ divided by the standard deviation $\sigma$ for the raw movie; the skewness from the background subtracted movie; an average fluctuation frequency $f_{av}$ obtained as a weighted average of the pixel-wise Fourier spectrum, $F(f,x,y)$, by the frequency range $f$ such that $f_{av}(x,y) = \sum_{f} fF(f,x,y) / \sum F_{f}(f,x,y)$; and a calculation of the apparent flow direction in the movie frame (described later).}
A rise in the \markchange{fluctuation amplitude} on a given pixel can be interpreted as an indication of a significant fluctuating component of the light picked up by that pixel. This naturally increases in regions where the camera sight line is near-tangent to the magnetic field \markchange{since more light is collected in this region per-fluctuation where line integration is maximised}. There is also a natural increase of the \markchange{fluctuation amplitude} in regions around the plasma-material interface, where local increases in the neutral density can lead to increased light emission. The \markchange{fluctuation amplitude} conforms well to the geometry of the scrape-off layer flux-surfaces in all four cases indicating that fluctuations observed in the movie can be associated with scrape-off layer fluctuations, commonly termed blobs or filaments \cite{D'IppolitoReview}. \markchange{Interestingly the region where the fluctuation amplitude peaks seems to remain similar in form in all four cases, independantly of the position of the secondary X-point. In addition the rise in fluctuation amplitude is echoed by a higher skewness.} There is a reduction in the \markchange{fluctuation amplitude and skewness, particularly notable at large $\rho_{X2}$,} in the region tending towards the primary X-point in each case. This is similar to observations made on MAST \cite{WalkdenNF2017} where the X-point was shown to be quiescent. 
\\\markchange{The average frequency of fluctuation, $f_{av}$, drops in regions where the fluctuation amplitude is high. This may indicate a tendency towards larger amplitude intermittent fluctuations in these regions, however it should be noted that such conclusions may be misleading since the fluctuations measured on a given pixel are not necesarilly spatially localised and are subject to line-integration effects. $f_{av}$ rises in the region below the primary X-point, with the strongest rise corresponding to the lowest value of $\rho_{X2}$.} In \markchange{the lowest row of} figure \ref{Fig:Fluct_amps_flow} the typical direction of the motion of structures in the movie is shown (given by the direction and magnitude of the arrows within the quiver diagram). This \markchange{pattern of motion} is extracted using a correlation analysis \markchange{where,} for each image pixel respectively, the \markchange{the motion is measured by calculating the spatial gradient (in the image plane) of the time derivative of the correlation function of that pixel with its nearest neighbours.} The full vector is calculated from the \markchange{x and y components of the spatial gradient}. \markchange{Labelling the time delay used in the correlation as $\delta t$ and the $x$ and $y$ indices as $i,j$, this can be written
\begin{equation}
    v_x \propto \left(C(\delta t,i+1,j) -  C(-\delta t,i+1,j)\right) - \left(C(\delta t,i-1,j) + C(-\delta t,i-1,j)\right) 
\end{equation}
and likewise for $v_{y}$, where $C$ is a 3x3 correlation matrix calculated by correlating a given pixel with its 8 nearest neighbours with a delay of $\delta t$. This method does not directly measure the flow velocity, so the information flow is not quantified, however it captures the qualitative features of the flow and in particular, its directionality. This works particularly well in regions of the image where filaments move as coherent objects. More advanced methods to calculate the information flow within the movies were tested, however these proved unreliable due to the limited spatial and temporal resolution of the movies under study. For ease of explanation, the term 'flow' will be used to describe the vector field shown in figure \ref{Fig:Fluct_amps_flow} with the discussion above serving as a caveat to this defintion.} There are two distinct flow patterns apparent in figure \ref{Fig:Fluct_amps_flow} \markchange{which have been separated out based on the direction of their vertical component (red for upwards flow, cyan for downwards)}. There is a flow that is diagonal on a bearing of roughly $45$ degrees (clockwise from \markchange{vertically upwards}) and can be seen in all four cases at the right hand side of the frame, \markchange{which exists in the region where the fluctuation amplitude peaks and the frequency lowers}. The direction of the flow points poloidally, roughly parallel with the magnetic surfaces local to that region of the image. It should be noted that this apparent poloidal motion cannot be disambiguated since a true poloidal flow or a poloidal projection of a toroidal flow of field-aligned structures will appear the same in the two-dimensional movie. Since there is no neutral beam heating in the plasmas measured however, it is likely that externally driven toroidal flows will be minimal. The observed poloidal flow points in the ion diamagnetic direction (noting that the x-axis is reversed in all figures from the camera image plane presented here) which is also the direction expected of the scrape-off layer ExB flow assuming the potential in the scrape-off layer is set by the sheath potential which increases \markchange{monotonically} towards the separatrix. This poloidal flow is present in all cases, though as $\rho_{X2}$ decreases, the region where the poloidal flow is present becomes more isolated to the OSOL. There is a second flow that becomes more prevalent as $\rho_{X2}$ decreases and exists in the region between the two nulls. The flow has a bearing that varies between roughly $100$ and $170$ degrees at differing points in the image plane and crosses magnetic flux surfaces, indicating that it \markchange{may contribute to cross-field transport}. The pattern of the flow is consistent with a flow across flux-surfaces in the ISOL region, that is projected along magnetic fieldlines onto the camera image plane. The increased prevalence of this radial flow as $\rho_{X2}$ increases implies that the additional null in the SF LFS- configuration \markchange{may be impacting} the properties of turbulence in the ISOL. Indeed in the case of 52103 the radial flow appears to extend across the second X-point, possibly providing a mechanism for transport into the private flux region connecting to SP3. \markchange{It should be noted that the change in the magnetic geometry from 52103 to 52113 moves SP2 down the centre column and lower in the camera frame. It is therefore possible that fluctuations exist close to the SP2 strike point which occupy only a small portion of the camera frame and therefore cannot be accounted for in this analysis. }
\\The analysis presented so far indicates \markchange{that} a deeper analysis of the fluctuation characteristics ISOL and OSOL may be useful in \markchange{understanding the complex dynamics observed in these movies}. In addition to the ISOL and OSOL regions, the region close to the primary X-point will also be studied in some detail to investigate whether the quiescence observed in figure \ref{Fig:Fluct_amps_flow} is borne out in a deeper analysis of the fluctuation structures observed. 
\subsection{Spatial characteristics of the fluctuations}
\label{Sec:Spatial}
In this section the spatial structure of fluctuations in each of the three analysis regions will be investigated. The analysis proceeds by taking a representative pixel from each region and calculating the instantaneous cross-correlation of the data on that pixel with all other pixels within the frame over the time-series. Through this method the typical structure in the image frame of a fluctuation that crosses the selected pixel is obtained. This method has been used for similar purposes in the MAST divertor \cite{WalkdenNF2017} where also the method was validated on synthetic data. Regions in the image showing a correlation of less that 15\% with the selected pixel are set to zero in order to isolate the structure of the fluctuations under study. \markchange{This method has been compared to a conditional averaging method which determines the average shape of fluctuations passing a given pixel above a fluctuation threshold. Results were entirely consistent between both methods.}
\\Figure \ref{Fig:Ccorr_1} shows the cross-correlation carried out in the OSOL, ISOL and X-Point regions of the image for each of the four plasmas in the $\rho_{X2}$ scan. Each region will now be analysed in turn:
\begin{figure}[htbp]
    \includegraphics[width=\textwidth]{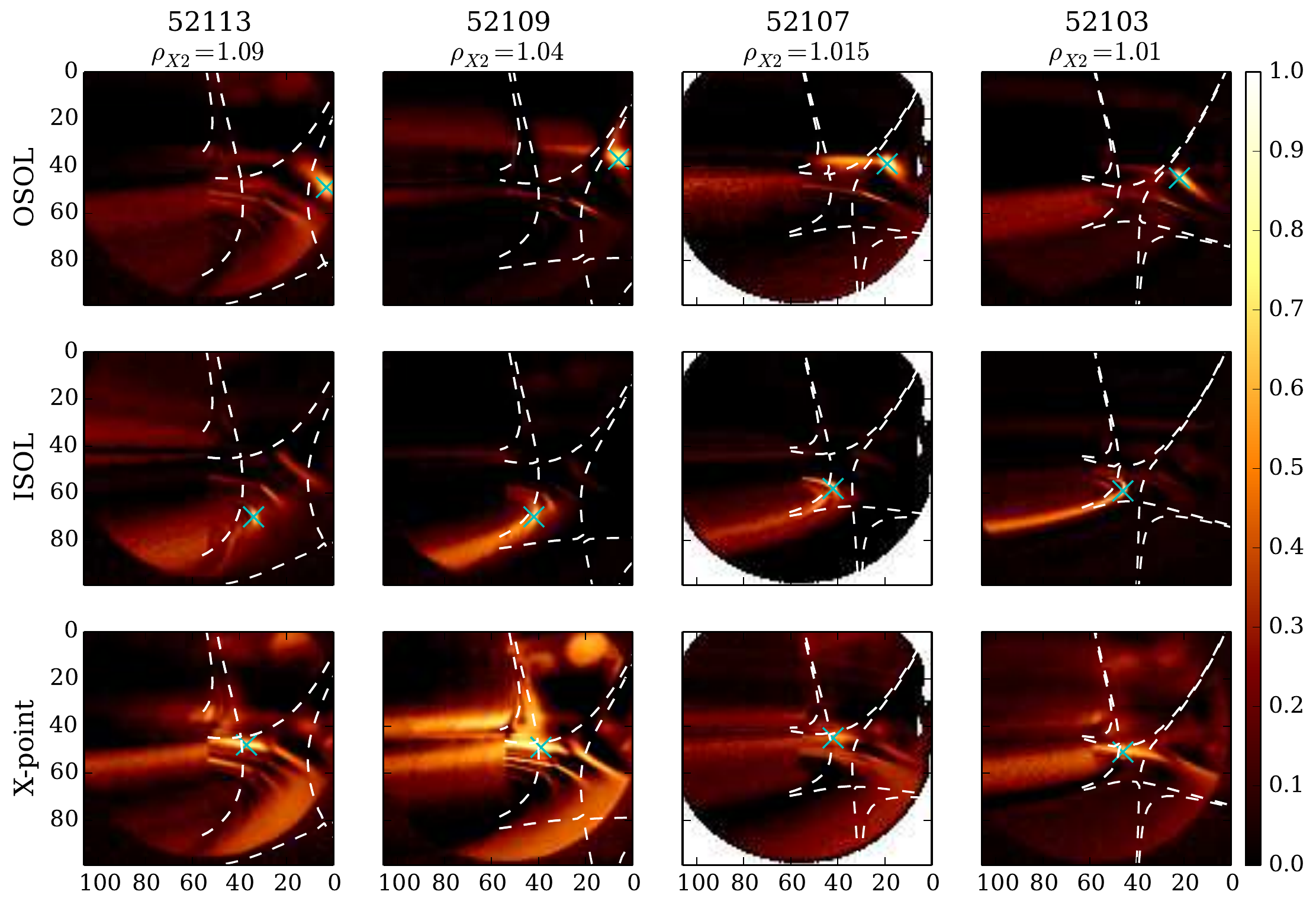}
    \caption{Cross-correlation analysis carried out for each plasma in the scan of $\rho_{X2}$ with representative pixels from the OSOL region (upper row), the ISOL region (middle row) and the X-point region (lower row).}
    \label{Fig:Ccorr_1}
\end{figure}
\subsubsection{OSOL}
\begin{figure}[htbp]
    \includegraphics[width=\textwidth]{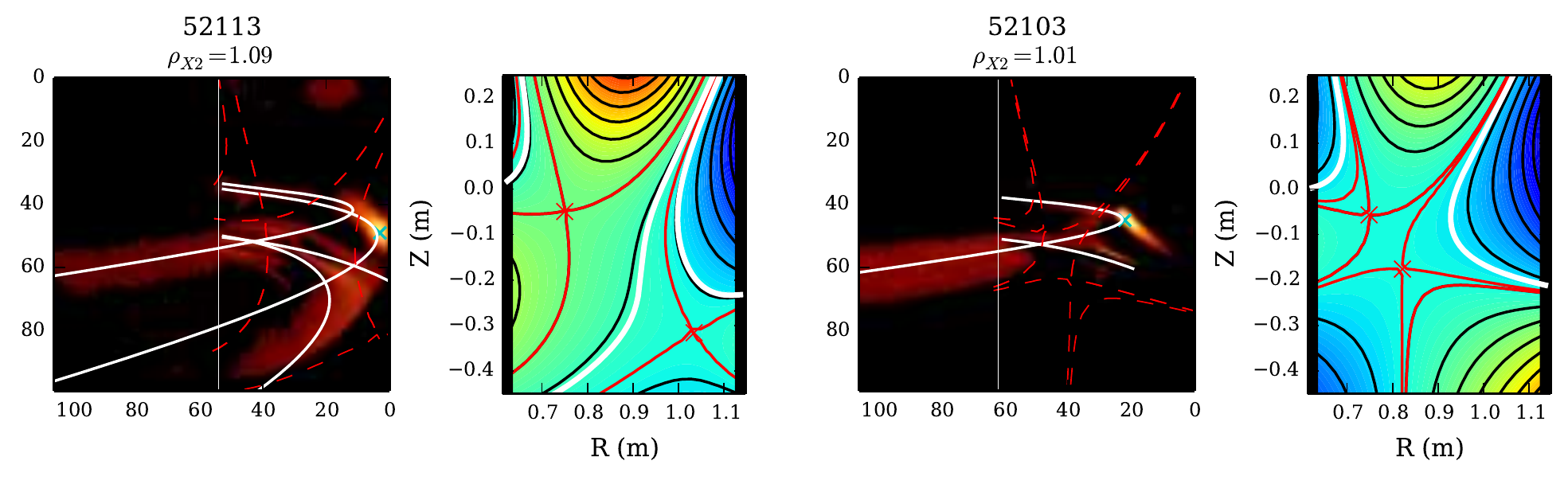}
    \caption{Magnetic field-lines projected onto the camera field of view that overlay the structures exhibited in the OSOL region of plasmas 52113 and 52103, the extrema of the $\rho_{X2}$ scan. Also shown are the corresponding magnetic flux surfaces that the projected field-lines lie on. This analysis is carried out qualitatively since no good metric has been found to assess the quality of the projected field-line. For 52113, two magnetic field-lines have been chosen which intersect the correlation region surrounding the second pixel inside and outside the second separatrix respectively.}
    \label{Fig:OSOL_FL_trace} 
\end{figure}

In the OSOL region the shape of the cross-correlation in the local vicinity of the selected pixel is similar in all four cases. The area surrounding the selected pixel is close the point where fluctuations that align with magnetic field-lines lie almost tangent to the camera LOS and so the emission that is captured primarily represents the cross-section of the fluctuation. In all four cases the fluctuations have a tendency towards an elliptically shaped cross-section with the dimension normal to the flux-surface larger than the dimension parallel to the flux surface. The ellipticity of the cross-section increases as $\rho_{X2}$ decreases, as might be expected from the increase in flux expansion approaching the null region when $\rho_{X2}$ is low. The elliptically shaped cross-section is consistent with these filaments originating further upstream and distorting in shape by following the topology of the magnetic field \cite{FarinaNF93,WalkdenPPCF2013,WalkdenThesis}. Furthermore each case shows regions of heightened correlation that are not directly connected to the selected pixel. This indicates a physical connectivity between the structure that overlaps the selected pixel and disconnected areas of the camera image plane. The most likely cause is alignment of the fluctuating structures to the background magnetic field. This can be demonstrated by projecting the trajectory of magnetic field-lines onto the camera image plane via a registration of the camera position using the \texttt{calcam} code \footnote{Available at https://github.com/euratom-software/calcam}. In figure \ref{Fig:OSOL_FL_trace} magnetic field-lines have been projected on the OSOL cross-correlation image that pass through the regions of high correlation for the two cases at the extreme ends of the $\rho_{X2}$ scan. \markchange{The position of the centre column of TCV has been indicated by a vertical line and only portions of the magnetic field-line that would be visible in the camera view (i.e not shadowed by the centre column or other machine components) are drawn.}
\\\markchange{Magnetic field-lines mapped onto the cross-correlation function are able to connect all regions of heightened correlation and} the cross-correlation structure shown in the OSOL is consistent with filamentary fluctuations maintaining a coherent structure along the magnetic field-line. Notably the structures in 52113 straddle the secondary separatrix, indicating that filaments from upstream are able to connect all the way through to SP2. This has been shown by projecting two magnetic field-lines that intersect the region of heightened correlation surrounding the selected pixel both inside and outside of the second separatrix. With both field-lines plotted, the correlation is relatively well mapped out. In the smaller $\rho_{x2}$ cases the correlation appears confined to the outer SOL connecting to SP4 and a second field-line inside the secondary separatrix is not required to describe the shape of the correlation. \markchange{In both cases the field lines map to the region in front of the centre column, where the foreground part of the filamentary structure can be seen. This correlation is highest when the centre column shadows the far-field plasma since, in regions where the far-field is visible, fluctuations in the far-field are decorrelated with the selected pixel, reducing the net correlation in that region. As such, discontinuity in the correlation function as it maps along the magnetic field line is a measurement artifact and does not indicate a physical discontinuity in the filamentary structure. }
\subsubsection{ISOL}
\begin{figure}[htbp]
    \includegraphics[width=\textwidth]{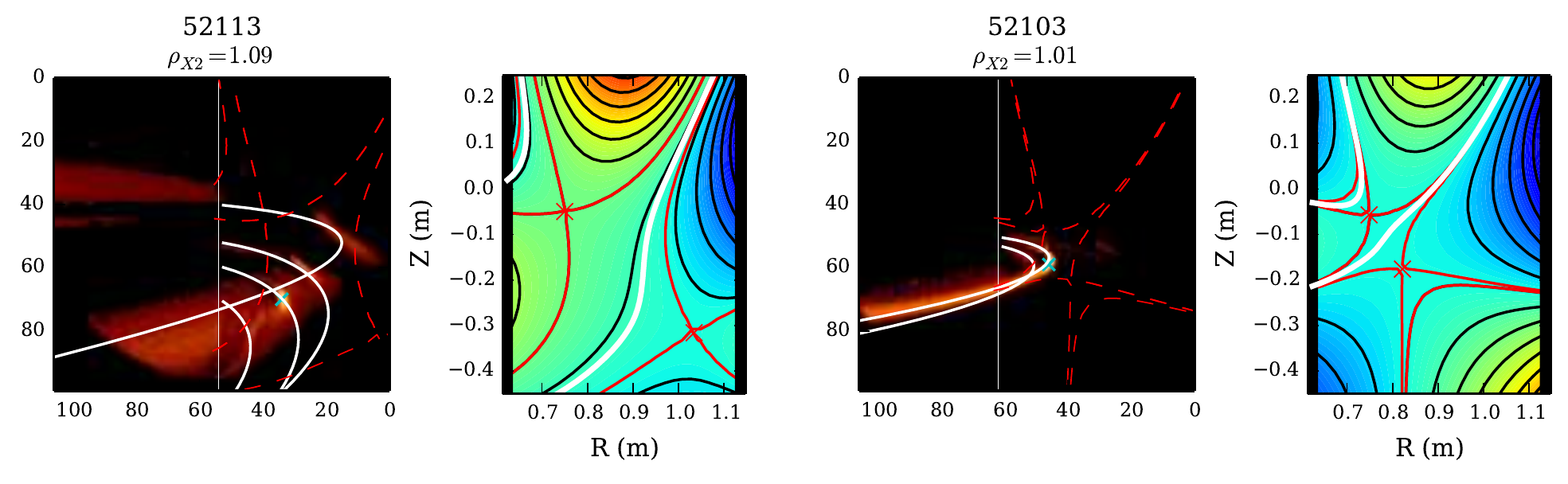}
    \caption{Similarly to figure \ref{Fig:OSOL_FL_trace}, magnetic field lines are projected such that they overlay regions of high correlation in the ISOL region.}
    \label{Fig:ISOL_FL_trace} 
\end{figure}

In the ISOL region a distinctly different behaviour is apparent as $\rho_{X2}$ decreases. At large $\rho_{X2}$ the structure of the correlation in figure \ref{Fig:Ccorr_1} in the ISOL is similar to the structure found by selecting a pixel in the OSOL region. It shows multiple regions of correlation corresponding to the fluctuation connecting along the magnetic field both upstream and towards the divertor, with an elliptically shaped cross-section. Since similar structures are evident whether a pixel is selected in the OSOL or ISOL region at large $\rho_{X2}$ it may be concluded that there is a true physical connection between the two regions. As in the OSOL case, this can be confirmed by projecting magnetic field-lines onto the correlation images. This is done for the two extreme cases of the scan in $\rho_{X2}$ in figure \ref{Fig:ISOL_FL_trace}. 
\\At large $\rho_{X2}$ the correlation follows the trajectory of a magnetic field-line well in both the direction upstream and downstream, indicating a connectivity of the filamentary structures between the upstream and downstram regions along the magnetic field line connecting to SP2. This is not apparent in the smaller $\rho_{X2}$ cases. At small $\rho_{X2}$ the filamentary structure does not correlate for more than a maximum of two turns around the machine, and does not correlate with the upstream plasma. In the ISOL region the poloidal field is low and the field-line wraps tightly around the machine. Despite this, the poloidal deviation of the fieldline after one turn of the machine is still significant enough that, were the filament to connect further upstream, it should be expected to be distinguishable. Furthermore the cross-sectional shape of the fluctuations in the small $\rho_{X2}$ case appear qualitatively less elliptical in the ISOL than the OSOL. Since the topological distortions of the filament cross-section due to the magnetic field should \emph{increase} with proximity to the separatrix, if ISOL filaments originated upstream their cross-sections should be highly sheared. Since this is not the case, and since the correlation appears to be confined to the divertor region the evidence gathered here suggests that at small $\rho_{X2}$ filaments are generated locally in the region between the two nulls in the TCV LFS SF- configuration.\markchange{It is once again worth noting that the camera viewing geometry may obscure fluctuations in 52113 that are particularly local to strike point SP2. If such fluctuations exists though, the analysis presented here suggests that they must be strongly localised to the strike-point, since no clear indication of their presence in the ISOL region is present.}
\subsubsection{X-Point}
\label{Sec:X-Point}
\begin{figure}[htbp]
    \includegraphics[width=\textwidth]{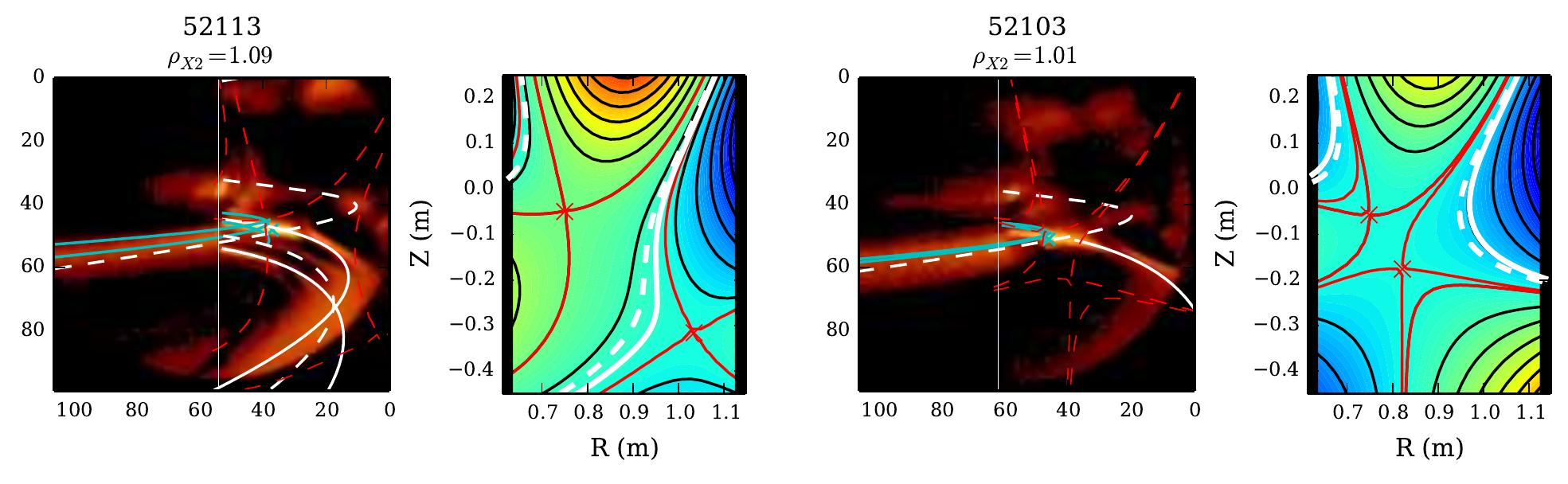}
    \caption{Similarly to figure \ref{Fig:OSOL_FL_trace} and \ref{Fig:ISOL_FL_trace}, magnetic field lines are projected such that they overlay regions of high correlation in the  region local to the inner X-point. In each case two separate magnetic field-lines are required to fully characterise the structure. These are represented by solid and broken white lines respectively. Cyan lines represent a projection of a magnetic fieldline situated close to the X-point twice around the machine.}
    \label{Fig:XP_FL_trace} 
\end{figure}

Figure \ref{Fig:Fluct_amps_flow} suggests that the primary X-point region in all four cases within the scan of $\rho_{X2}$ remains quiescent in the manner observed on MAST \cite{WalkdenNF2017}. Despite this apparent quiescence, the correlation analysis produces rather complex structures when a pixel is selected close to the inner X-point. To elucidate the nature of these structures, once again magnetic field-lines have been projected onto the correlation images for the two extreme cases in the $\rho_{X2}$ scan, shown in figure \ref{Fig:XP_FL_trace}.
\\In both cases two magnetic field-lines have been identified (one as a solid line and the other as a broken line) which map onto the regions of high correlation. The field lines both lie on flux-surfaces that are a significant distance into the SOL, away from the primary separatrix which intersects the primary X-point. This shows that the structures present can be described as filaments in the OSOL region that cross the X-point region in the image frame due to the projection of their toroidal structure onto the camera view. These structures are not located near the X-point in the poloidal plane but provide a source of emission that crosses the pixel selected in the X-point region in the image frame. Since the correlation structures local to the X-point region can be fully explained by these filaments that exist far from the X-point, it can be concluded that, as with the MAST analysis, the inner X-point of the TCV LFS SF- configuration is quiescent when measured with the tangential viewing unfiltered fast-camera at all values of $\rho_{X2}$ investigated. For comparison, fieldlines that are in close proximity to the X-point have been projected onto the camera view (cyan lines in figure \ref{Fig:XP_FL_trace}). Being close to the null, these fieldlines wrap very tightly around the machine. Whilst they may partially align with the correlation structure as they wrap around towards the camera, in front of the centre column, they cannot explain the complex structure of the correlation elsewhere. This, coupled with the observation from figure \ref{Fig:Fluct_amps_flow} that the fluctuation amplitude, skewness and apparent flow drop in the vicinity of the primary X-point, supports the assertion that the primary X-point appears quiescent. It is also worth noting that in 52113 the QXR terminates in the SOL well inside of the secondary separatrix, indicating that the presence of the QXR is not reliant on the presence of the secondary null point. This suggests that single null plasmas in TCV should be expected to show similar behaviour around the X-point. 
\subsection{Temporal characteristics of the fluctuations}
\begin{figure}[htbp]
    \includegraphics[width = \textwidth]{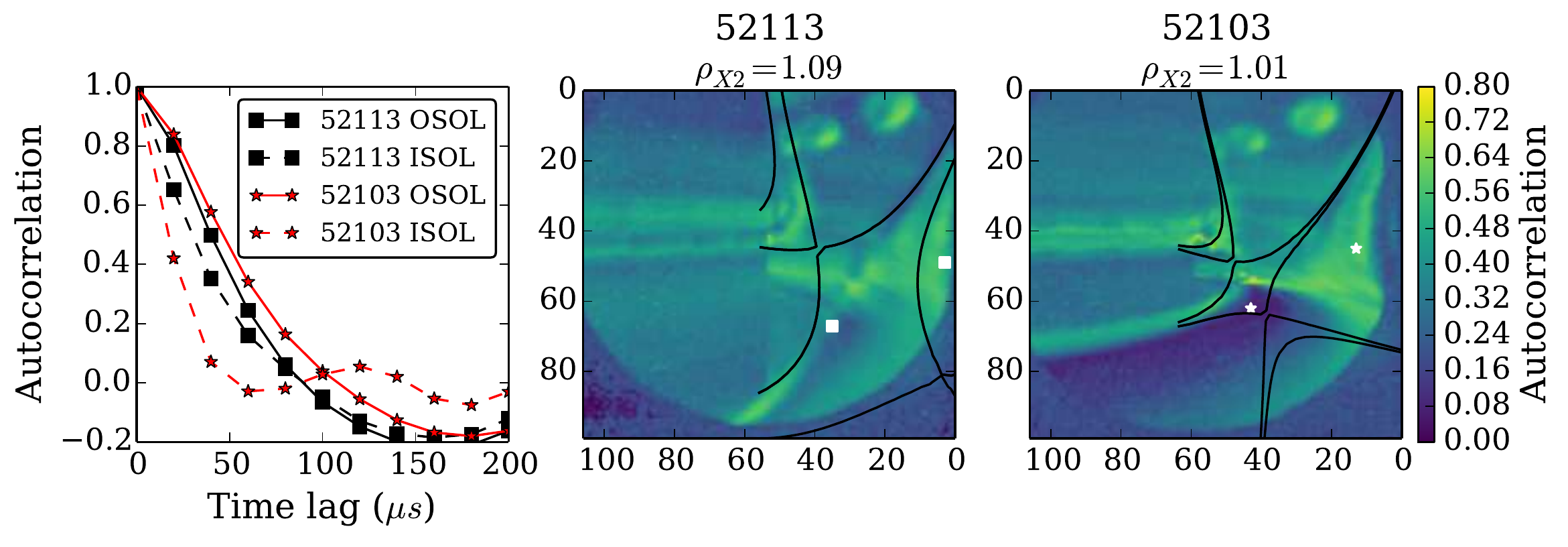}
    \caption{Left: Autocorrelation function in the ISOL and OSOL (positions marked on central and right hand figures) for plasmas 52103 and 52113 respectively. Centre and right: Pixel-wise autocorrelation calculated at a time-lag of two frames ($40\mu s$) for the image time-series in 52113 and 52103.}
    \label{Fig:Acorr}
\end{figure}

\begin{figure}[htbp]
    \includegraphics[width=\textwidth]{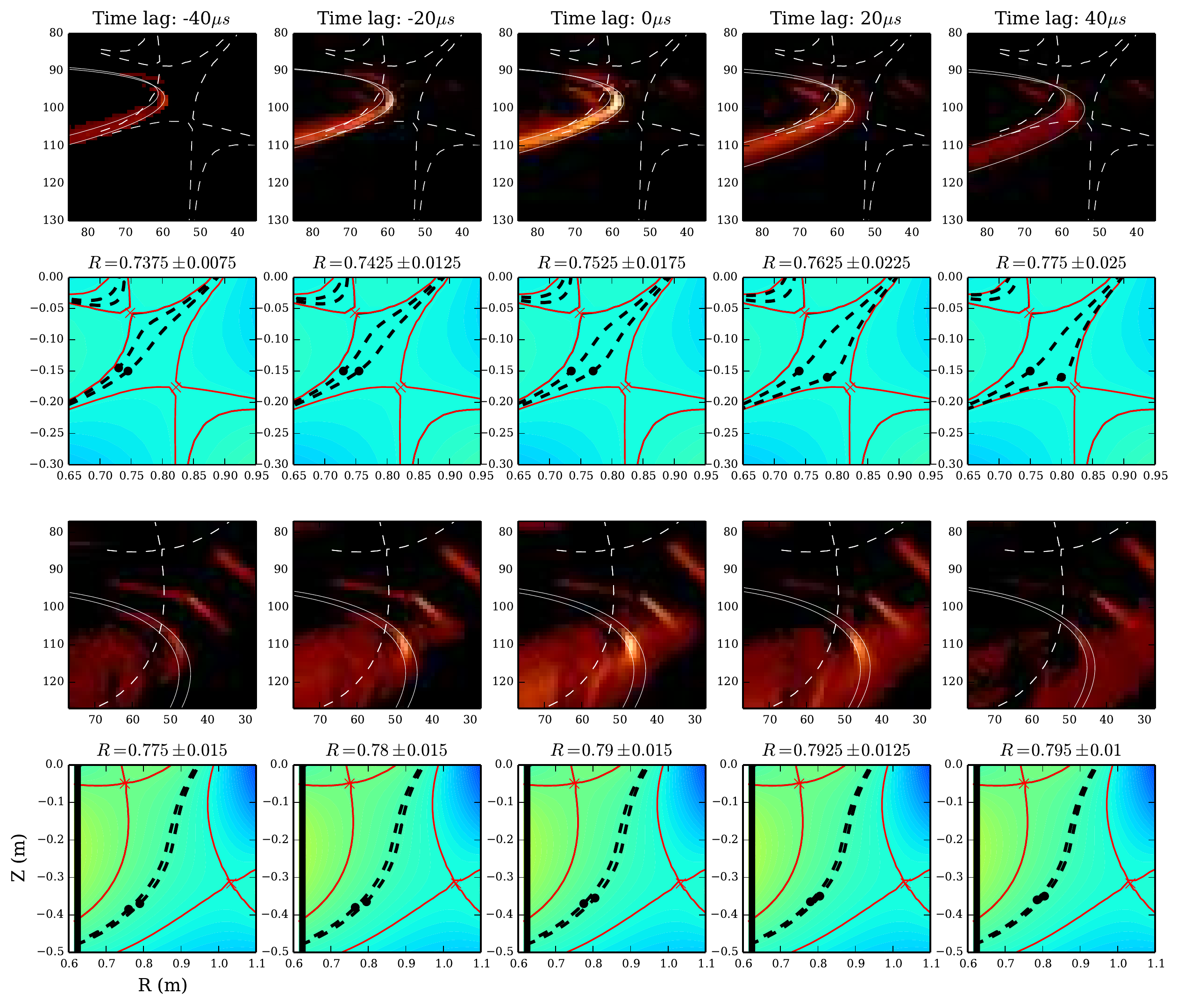}
    \caption{Time-delayed cross-correlation in the ISOL regio stretching from a delay of two frames in the past to two frames in the future. Magnetic field lines have been projected which qualitatively encompasses the region of heightened correlation. This region is tracked through the series. Magnetic flux surfaces are indicated with dashed lines and positions in the poloidal plane at the camera tangency angle, indicated as closed circles, of the two magnetic field-lines projected in the camera image. The upper two rows correspond to plasma 52103 ($\rho_{X2} = 1.01$) whilst the lower two rows correspond to plasma 52113 ($\rho_{X2} = 1.09$). }
    \label{Fig:ISOL_fl_track}
\end{figure}

\begin{figure}[htbp]
    \includegraphics[width=0.8\textwidth]{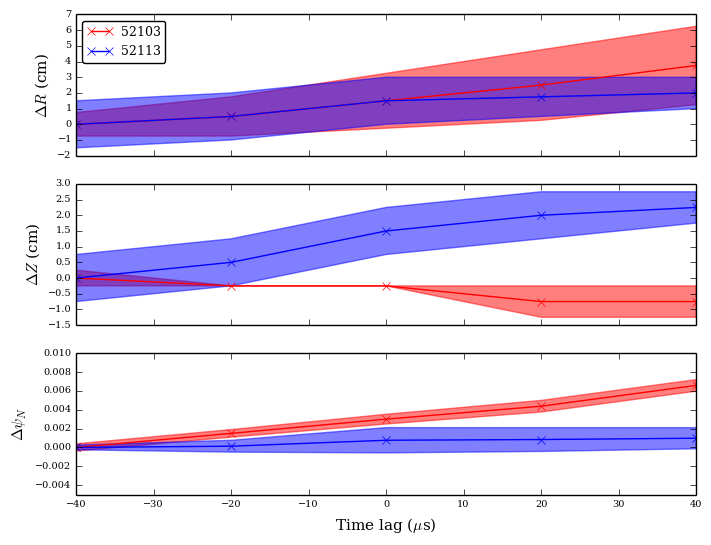}
    \caption{ Radial (upper), vertical (middle) and normalised poloidal flux (lower) trajectories of the ISOL structures in tracked in figure \ref{Fig:ISOL_fl_track} in the poloidal plane. The shaded region corresponds to the region enclosed by the field-lines used to track the structures in figure \ref{Fig:ISOL_fl_track}. Red curves correspond to plasma 52103 whilst blue curves correspond to 52113.}
    \label{Fig:Trajects}
\end{figure}

The spatial structure of fluctuations in the three regions defined in the previous section, OSOL, ISOL and XP, were shown to vary as $\rho_{X2}$ varied. The observation that, as $\rho_{X2}$ decreases, filaments in the ISOL region become de-correlated from upstream, may also indicate that their motion will vary in comparison to the OSOL region. To investigate the temporal characteristics of the fluctuations a time delay can be introduced into the correlation. As a general quantifier of the temporal characteristics of fluctuations, the pixel-wise auto-correlation has been calculated at a time-lag of two frames. This can be interpreted as showing how strongly decorrelated a pixel becomes two frames after a fluctuation is present. Carrying this out for each pixel in the camera view shows regions of the image where pixels become decorrelated more quickly, indicating more rapid fluctuations. Figure \ref{Fig:Acorr} shows this time-lagged pixel-wise autocorrelation at $\rho_{X2}=1.09$ and $\rho_{X2} = 1.01$ respectively. Also shown is the autocorrelation function for both cases sampled in the OSOL and ISOL respectively. 
\\The region that has been associated with the localised ISOL filaments at smaller $\rho_{X2}$ shows a pronounced reduction in the time-lagged autocorrelation compared with the OSOL region and with both regions at larger $\rho_{X2}$\markchange{, consistent with the measurement of $f_{av}$ in figure \ref{Fig:Fluct_amps_flow}}. In the ISOL the autocorrelation function contracts as $\rho_{X2}$ decreases. This indicates that fluctuations in the ISOL region at small $\rho_{X2}$ may evolve faster than their counterparts in the OSOL. Changes to the temporal properties of the filaments in the ISOL as $\rho_{X2}$ decreases may indicate changes in their propagation and consequently changes to the transport that can be associated with them. \markchange{Since these measurements are made in the lab frame, changes in the bulk rotation in different regions of the plasma, or in diffent plasmas may affect the autocorrelation function. Therefore, it is not apriori clear that the drop in autocorrelation function in the ISOL region indicates a change in filamentary dynamics. To investigate this further,} the motion of fluctuations in the ISOL can be tracked by introducing a time delay into the cross-correlation analysis introduced in section \ref{Sec:Spatial}. At a positive (negative) time-delay, the cross-correlation presents the typical structure found on the camera in the future (past) after a fluctuation is measured on the selected pixel. In this case time-delays are introduced stretching from two frames in the past to two frames in the future, giving a total window of 80$\mu s$. Magnetic field-line projections are again used to infer the spatial structure of the cross-correlation, however this time two magnetic field-lines are projected that qualitatively bound the region of heightened correlation. The flux-surfaces that these magnetic field-lines lie on, as well as their position in the poloidal plane at the camera tangency angle, are shown in figure \ref{Fig:ISOL_fl_track}. 
\\In both cases the ISOL fluctuations evolve, however the details of their propagation differ. In 52103 the motion is mainly in the radial direction and the region of heightened correlation expands radially. By contrast in 52113 the radial motion and radial expansion is suppressed compared to 52103. There is a slightly more pronounced vertical motion in 52113, though this is difficult to distinguish visually. A more quantitative comparison is given in figure \ref{Fig:Trajects} where the poloidal plane trajectories of the structures in figure \ref{Fig:ISOL_fl_track} are shown as a function of time. 
\\Figure \ref{Fig:Trajects} shows a moderately faster radial motion in 52103 compared to 52113, but a reduced vertical motion. When mapped into the normalised poloidal flux ($\psi_{N}$) the difference between the two types of motion becomes clear. In 52103, where $\rho_{X2} = 1.01$ and the nulls are close, the motion is predominately across the magnetic field such the the fluctuations move outwards in $\psi_{N}$ towards the secondary separatrix. By contrast the motion observed in 52113 remains approximately stationary in $\psi_{N}$ indicating a predominantly poloidal flow. It can therefore be concluded that as the nulls of the TCV LFS SF- contract towards smaller $\rho_{X2}$ isolated fluctuations in the ISOL region start to develop which provide a intermittent cross-field flux that enhances transport between the nulls.
\section{Discussion}
\label{Sec:Disc}
The main observation made here is that as $\rho_{X2}$ decreases in the LFS SF- configuration radial transport, mediated by intermittent filamentary fluctuations, increases in the region between the two null points due to the localised production of filaments that propagate radially between the nulls. This observation is supported by recent measurements on TCV using both infra-red and probe diagnostics that show an enhanced level of transport resulting in broader profiles between the nulls of the TCV LFS SF- configuration compared to the single-null case \cite{RobertoCommun}. The LFS SF- configuration has also been shown to support a radiation front away from the primary X-point in the ISOL region in conditions where the single-null case radiates from the X-point and inner divertor leg \cite{ReimerdesNF2017}. It is possible that enhanced filamentary transport in the ISOL region may play a role in broadening profiles within the null region and helping to move the radiation front away from primary null. This paper has also demonstrated the presence of a quiescent region near the primary X-point (QXR), a phenomenon first observed in MAST. The interplay between the QXR and the fluctuations observed in the OSOL and ISOL is difficult to determine from the camera footage, though it may be notable that the ISOL localised fluctuations appear most strongly when the secondary X-point lies on a flux that coinsides with the QXR. It is likely that detailed non-linear numerical simulation will be required before the underlying physical mechanisms governing the OSOL and ISOL fluctuations and the QXR can be understood.
\\One notable difference between the fluctuations in the large and small $\rho_{X2}$ cases is the the connectivity of filaments along magnetic field-lines. In particular the ISOL localised filaments observed in the small $\rho_{X2}$ case appear to correlate strongly at most twice toroidally along the magnetic field. In the ISOL region the poloidal magnetic field is substantially reduced by virtue of the proximity to the nulls, and therefore the magnetic field followed by the filaments is dominantly toroidal (more so than in the outer region of the SOL, or in the case when $\rho_{X2}$ is large). In this scenario a model was proposed by Ricci and Rogers \cite{RicciPRL2010} following measurements on TORPEX \cite{MullerPPCF2009} whereby a filament at its development phase is able to overlap itself and consequently short-circuits the current paths that determine its motion. It is not clear whether this model is consistent with filaments in the SF LFS- ISOL, however given that these filaments appear to form in a region of much lower poloidal magnetic field than their upstream counterparts, it is reasonable to consider whether differing mechanisms for current closure may apply. In addition filament motion in the vicinity of an X-point has been studied on TORPEX \cite{AvinoPRL2016} and numerically \cite{ShanahanPPCF2016} which suggested that background flows in the vicinity of the X-point may affect the trajectory of filaments. 
\\This study has only considered the SF LFS- configuration, however as discussed in the introduction and in refs \cite{LabitNME2017,VijversNF2014}, other topological forms of the snowflake divertor exist. In the SF+ configuration the second X-point lies in the private-flux region of the first, meaning that it cannot lie within the QXR of the outboard SOL. Likewise in the HFS SF- configuration the second X-point lies within the inboard SOL, so once again does not overlap with the QXR in the outboard SOL. It is unclear how the results of this study transfer to these other two configurations and this should be pursued to provide a fuller understanding both of the fluctuation characteristics of snowflake divertors and the role played by the null region on turbulent fluctuations. A study of the density dependance of the ISOL and OSOL fluctuations would also be a good avenue of future work.

\section{Conclusions}
\label{Sec:Conc}
This contribution analyses fast visibile imaging footage from a tangentially viewing unfiltered camera of the null region in the TCV low-field side snowflake minus (LFS SF-) configuration. Four plasmas are compared that comprise a scan in the quantity $\rho_{X2}$ which parameterises the distance between the two X-points of the LFS SF- configuration. As $\rho_{X2}$ decreases and the X-point gap contracts the fluctuation amplitude picked up by the camera becomes increasingly peaked in the outboard SOL region. In addition two distinguishable apparent flow patterns are present in the movies. The first corresponds to a poloidal flow of fluctuations in the outer SOL region (OSOL)  and is prominent in all plasmas. The second corresponds to a radial motion in the inner SOL (ISOL) region between the two null points and becomes increasingly prominent as the X-points contract towards one another. Based on this observation, a cross-correlation technique was used to analyse the spatial structure of fluctuations in both the OSOL and ISOL regions, as well as locally around the X-point. The primary X-point is shown to be quiescent, consistent with a previous study of the X-point region of MAST. At larger $\rho_{X2}$ fluctuations are observed to connect along magnetic field-lines between upstream and downstream, into the ISOL region. At smaller $\rho_{X2}$ fluctuations in the divertor are uncorrelated with upstream, indicating a local production of turbulent structures between the two nulls of the LFS SF-. The typical motion of the ISOL structures is tracked and a distinct difference in the motion of filaments at large and small $\rho_{X2}$ is shown. At large $\rho_{X2}$ filaments in the ISOL region move predominantly in the poloidal direction along flux-surfaces. By contrast at small $\rho_{X2}$ ISOL filaments move mainly in the radial direction across flux-surfaces, indicating an enhancement to cross-field transport between the two nulls of the TCV LFS SF- configuration when $\rho_{X2}$ is small.

\section{Acknowledgements}
This work has been carried out within the framework of the EUROfusion Consortium and has received funding from the Euratom research and training programme 2014-2018 under grant agreement No 633053. The views and opinions expressed herein do not necessarily reflect those of the European Commission.

\section{References}
\bibliographystyle{prsty}
\bibliography{../Bibliography}

\end{document}